\documentclass{webofc}
\usepackage[varg]{txfonts}   
\usepackage{subfigure}
\usepackage[switch]{lineno}
\newcommand{\setParDis}{\setlength {\parskip} {0.006cm}}
\begin{document}
\title{Search for the Chiral Magnetic Wave in Pb--Pb collisions using the ALICE detector}
%
%

\author{\firstname{\rm Wenya} \lastname{Wu (for the ALICE Collaboration)}\inst{1}\fnsep\thanks{\email{wenya.wu@cern.ch}}
}

\institute{Key Laboratory of Nuclear Physics and Ion-beam Application (MOE), Institute of Modern Physics, Fudan University, Shanghai 200433, China}

\abstract{%
    The interplay of the chiral anomaly and the strong magnetic field ($\sim$ 10$^{15}$ T) created in heavy-ion collisions could give rise to a collective excitation in the quark--gluon plasma called the Chiral Magnetic Wave (CMW). This effect can be experimentally sought by the charge asymmetry ($A_{\rm ch}$) dependence of elliptic flow $v_{2}$. However, non-CMW mechanisms such as local charge conservation (LCC) coupled with collective flow can also lead to a similar dependence of $v_{2}$ on $A_{\rm ch}$. The triangular flow ($v_{3}$) measurement serves as a reference as it is not expected to be affected by the CMW. The $v_{2}$ and $v_{3}$ of charged hadrons as a function of $A_{\rm ch}$ measured in Pb--Pb collisions at $\sqrt{s_{\rm NN}}$ = 2.76 and 5.02 TeV are presented. In addition, the event-shape engineering (ESE) technique is adopted for the first time to quantitatively disentangle the CMW signal and the LCC background. The results indicate that the background effects dominate the CMW measurements.
}
\maketitle
\section{Introduction}
\setParDis
    \label{intro}
    Vacuum topological fluctuation in the hot and dense matter created in heavy-ion collisions causes the chiral anomalies of quarks, which can be regarded as local parity violation in strong interactions. The interplay of the chiral anomaly and the strong external magnetic field is expected to produce various chiral anomalous phenomena~\cite{RefTho1}. One phenomenon, called the chiral magnetic wave (CMW), could lead to wave propagation of electric charge resulting in a charged quadrupole configuration in the dense matter~\cite{RefTho2}.
    
    The difference of the elliptic flow ($v_{2}$) between positively and negatively charged particles is expected to depend linearly on the charge asymmetry ($A_{\rm ch}$) for the measurement of the electric quadrupole moment from the CMW~\cite{RefTho2}, i.e., $\Delta \large v_{2}  \equiv  v_{2}^{-} - v_{2}^{+} \approx r A_{\rm ch}$. Here the $A_{\rm ch}=(N^{-}-N^{+})/(N^{-}+N^{+}) $ denotes the average net charge in an event. The $v_{2}^{+}$ ($v_{2}^{-}$) represents elliptic flow of positively (negatively) charged particles and the $r$ is the extracted slope parameter. In general, slopes are estimated from $\Delta v_{2}$ normalized to $v_{2}$ (i.e. $r^{\rm Norm}$) to probe the intensity of CMW signals in different collision systems. In addition to the slope, there is another observable proposed in Ref.~\cite{RefIntCov} for studying the CMW. It is defined as the integral covariance ($\rm Int. \, Cov.$) between $v_{2}^{\pm}$ and $A_{\rm ch}$,
    \begin{equation}
    \setlength {\abovedisplayskip} {0.006cm}
    \setlength {\belowdisplayskip} {0.03cm}
            {\rm Int. \, Cov.} \equiv \left\langle v_{2}^{\pm} A_{\rm ch}\right\rangle-\langle A_{\rm ch}\rangle\left\langle v_{2}^{\pm}\right\rangle \approx \mp r\left(\left\langle A_{\rm ch}^{2}\right\rangle-\langle A_{\rm ch}\rangle^{2}\right) / 2=\mp r \sigma_{A_{\rm ch}}^{2} / 2.
    \end{equation}
    A key advantage of $\rm Int. \, Cov.$ is that it does not require correction for detector acceptance or experimental tracking efficiency. The use of $\rm Int. \, Cov.$ also avoids the reduction of the data sample from dividing events into different bins of $A_{\rm ch}$ in the slope measurement. In the CMW measurement, the difference of $\rm Int. \, Cov.$ between positively and negatively charged particles ($\Delta \rm Int. \, Cov.$) is applied.
    
    In previous analyses~\cite{RefExp1, RefExp2}, magnitudes of the slopes $r_{\Delta v_2}$ are consistent with theoretical predictions in semicentral collisions, which seems to confirm the CMW. On the other hand, theoretical calculations predict that the quadrupole moment is distorted by the third-order expansion of collision geometry, which would lead to a zero-magnitude slope of $v_{3}$~\cite{RefTho2}. However, the similar values of $r_{\Delta v_{2}}$ and $r_{\Delta v_{3}}$~\cite{RefExp3} imply that there are backgrounds contributing to the CMW measurement. In fact, local charge conservation (LCC) is widely regarded as the most important background ~\cite{RefLCC1}. This is further supported by the calculation from Ref.~\cite{RefLCC2} which shows that $r_{\Delta v_{2}}$ stems from the intrinsic property of $A_{\rm ch}$ in the LCC scenario. The observed $A_{\rm ch}$ would automatically select particles from different pseudorapidity ($\eta$) and transverse momentum ($p_{\rm T}$) ranges leading to various kinds of dependence such as $A_{\rm ch}-\Delta v_{2}$, $A_{\rm ch}-\Delta p_{\rm T}$, and $A_{\rm ch}-\Delta |\eta|$~\cite{RefLCC2}. Note that the slopes between $A_{\rm ch}$ and $v_{2}$ cannot disentangle the CMW signal and the LCC background directly. Recently, an approach to quantitatively disentangle the CMW signal and the LCC background using the event-shape engineering (ESE) method~\cite{RefESE} was introduced in Ref.~\cite{RefLCC3}. It was shown that the $\Delta \rm Int. \, Cov.$ is independent on $v_{2}$ in the pure CMW scenario, while in the background, it is proportional to $v_{2}$ for shape-selected events. In fact, ESE technique was successfully used in constraining the fraction of the Chiral Magnetic Effect signal~\cite{RefExp4}. 
\section{Analysis details}
\setParDis
\label{sec-1}
    The analysis is performed using Pb--Pb collisions at $\sqrt{s_{\rm NN}}=5.02$ TeV recorded in 2015 and 2018 by ALICE. Minimum bias trigger criteria, as well as central and semicentral trigger conditions, are required to define the data samples. Events with $z-$component of the reconstructed primary vertex ($V_{z}$) within $\pm10 \, \rm cm$ are selected. The collision centrality is estimated from the amplitude of the signal measured by the V0 detector which consists of the V0A $(2.8<\eta<5.1)$ and V0C $(-3.7<\eta<-1.7)$ sub-detectors. Inclusive charged particles are reconstructed using the Time Projection Chamber (TPC) and the Inner Tracking System (ITS) with kinematic ranges of $|\eta|<0.8$ and $0.2<p_{\rm T}<2.0$ GeV/$c$. The results are corrected for the azimuthal acceptance of the TPC and for the $p_{\rm T}$ dependence of the tracking efficiency.
    
    In this analysis, the two-particle Q-Cumulant method~\cite{RefFlow} is adopted to measure $v_{2}$ and $v_{3}$. This method expresses particle cumulants in the flow vectors (Q-vectors) with respect to the harmonic of order n, such as $Q_{n} = \sum_{i=0}^{M} e^{in\varphi_{i}}$, where M is the number of particles. In order to reduce the non-flow effect from short-range correlations such as resonance decay, jets, and quantum correlation, events are divided into two sub-events A ($-0.8<\eta<-0.3$) and B ($0.3<\eta<0.8$). 
    
    The ESE technique is employed to disentangle LCC background contributions from the potential CMW signal as proposed in Ref.~\cite{RefLCC3}. The event shape selection is performed based on the reduced flow vector given as $\large q_{n}=\frac{Q_{n}}{\sqrt{M}}$, where $M$ is the multiplicity and $Q_{n}$ is the magnitude of the flow vector. In this analysis, the $\rm Int. \, Cov.$ and $v_{2}$ are measured in ten classes of $q_{2}$ determined by the V0C detector. Note that the first and the last intervals of $q_{2}$ are discarded to remove impurities from neighboring classes.
\section{Results}
\setParDis
\label{sec-2}
\subsection{Slope parameters $r_{\Delta v_2}^{\rm Norm}$ and $r_{\Delta v_3}^{\rm Norm}$}
\begin{figure}[h!]
    \setlength{\abovecaptionskip}{0.00001cm}
    \setlength{\belowcaptionskip}{-0.5 cm}
        \centering
        \begin{subfigure}
            \centering
            \includegraphics[width=.4\linewidth]{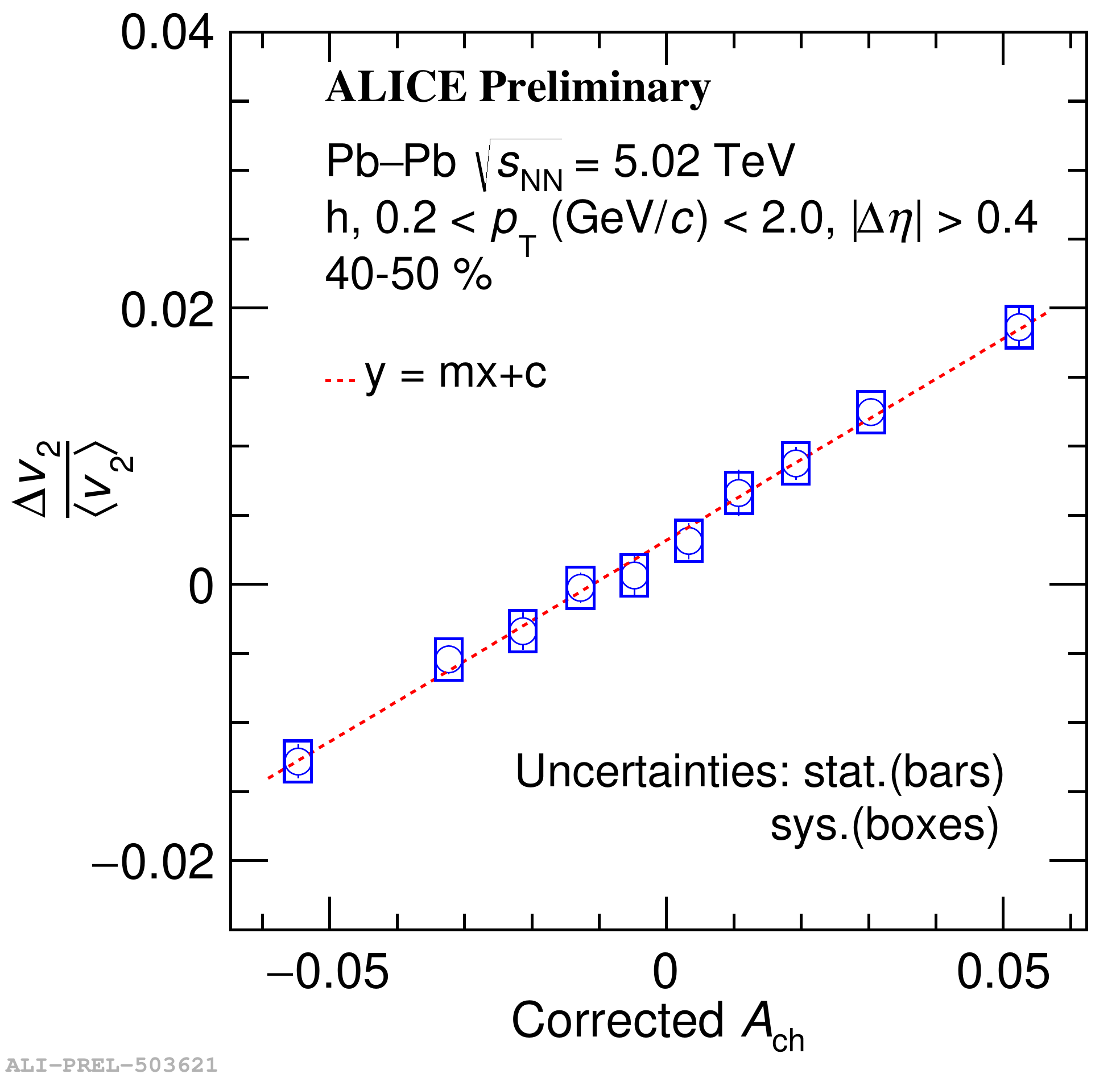}
        \end{subfigure}
        \begin{subfigure}
            \centering
            \includegraphics[width=.4\linewidth]{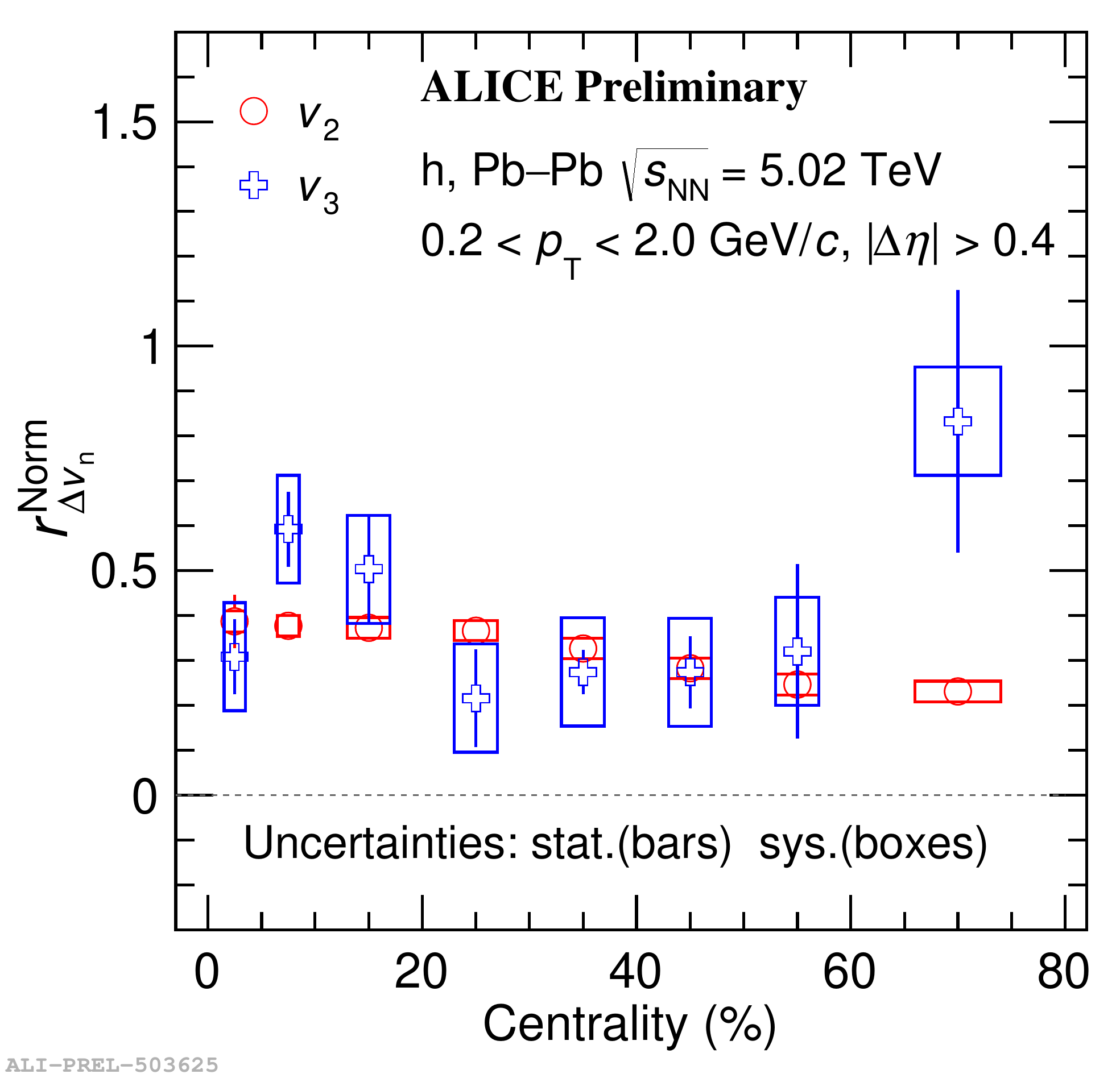}
        \end{subfigure}
    \caption{Left: normalized $\Delta v_{2}$ as a function of corrected $A_{\rm ch}$ for the 40--50\% centrality interval, with the fit function represented by a dash line. Right: slope parameter $r_{\Delta v_n}^{\rm Norm}$ for $v_{2}$ (red circles) and $v_{3}$ (blue crosses), after efficiency correction, as a function of centrality.} 
    \label{fig:sloppara}
    \end{figure}

    The left panel of Fig. \ref{fig:sloppara} shows the linear dependence of normalized $\Delta v_{2}$ on observed $A_{\rm ch}$ in the 40--50\% centrality interval, while the right panel shows the extracted the $r_{\Delta v_{2}}^{\rm Norm}$ (red circles) and $r_{\Delta v_{3}}^{\rm Norm}$ (blue crosses) after the efficiency correction. Clearly visible is a consistency between the second- and third-order results for most centrality intervals, confirming earlier measurements in Ref.~\cite{RefExp3}. The comparable magnitudes of $r_{\Delta v_2}^{\rm Norm}$ and $r_{\Delta v_3}^{\rm Norm}$ indicate that the background, which can be qualitatively explained by the LCC effect, dominates the measurements. 
\subsection{Fraction of CMW with ESE}
\begin{figure}[h!]
    \setlength{\abovecaptionskip}{0.0005 cm}
    \setlength{\belowcaptionskip}{-0.5 cm}
        \centering
        \begin{subfigure}
            \centering
            \includegraphics[width=.45\linewidth]{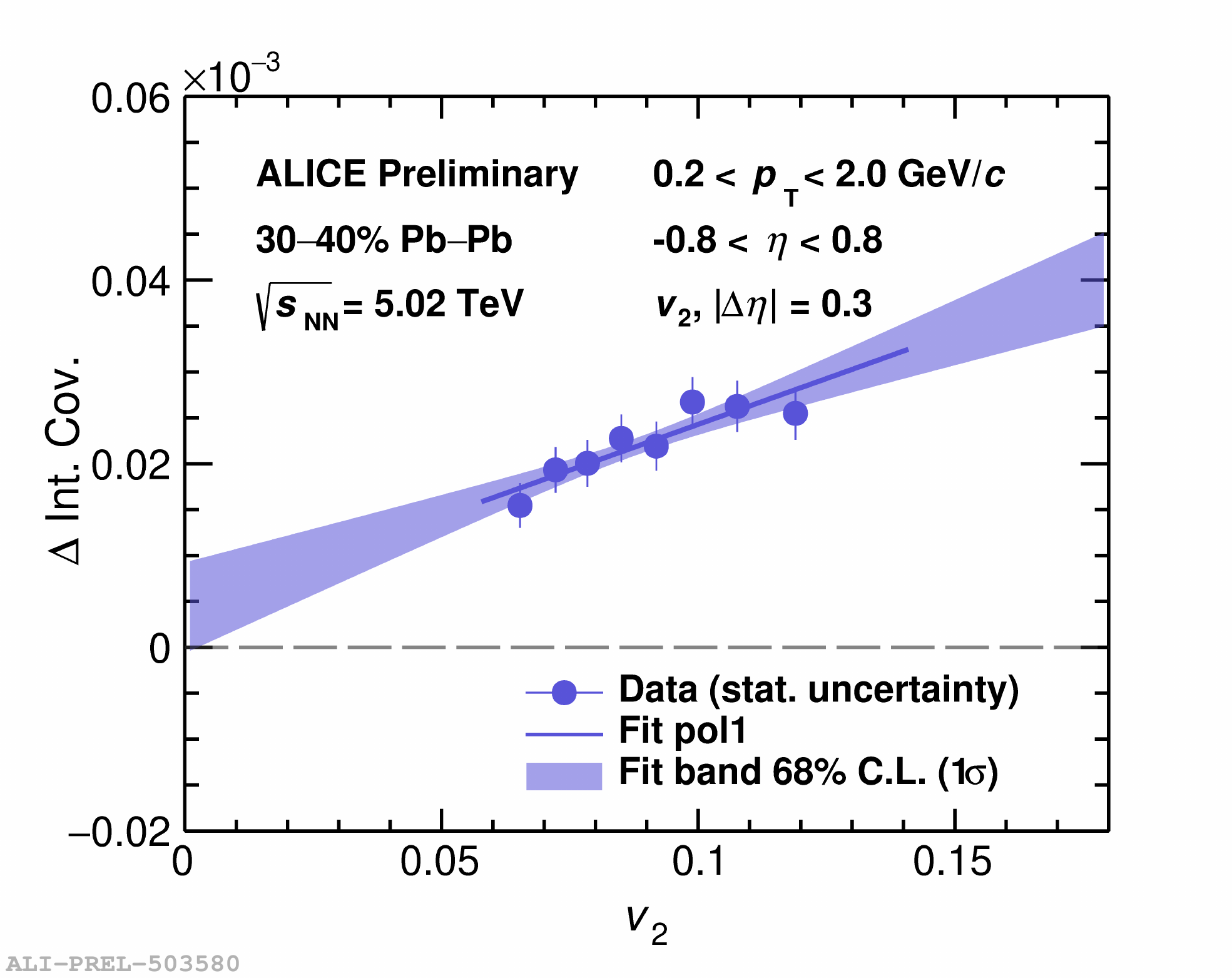}
        \end{subfigure}
        \begin{subfigure}
            \centering
            \includegraphics[width=.45\linewidth]{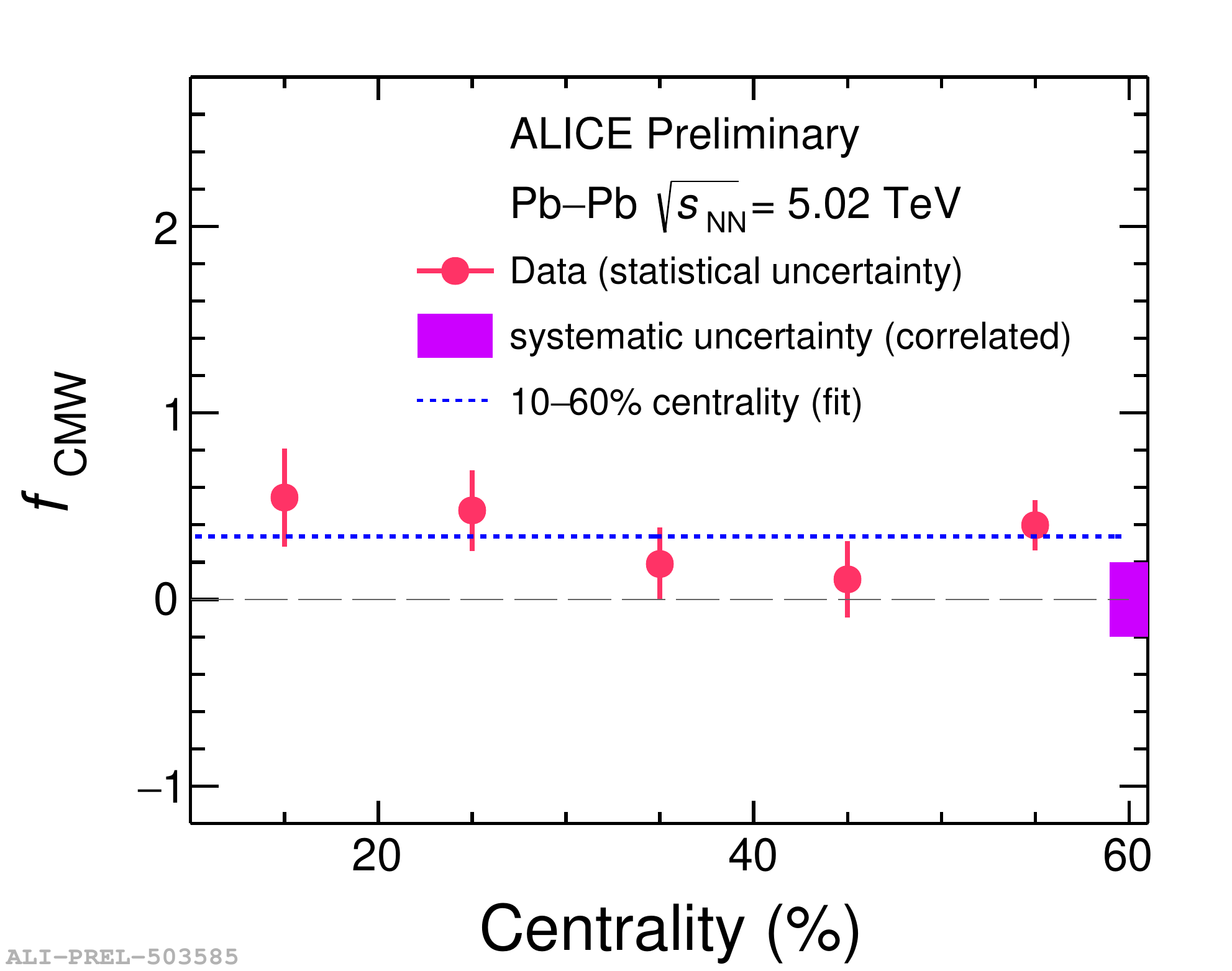}
        \end{subfigure}
    \caption{Left: Linear dependence of $\Delta \rm Int. \, Cov.$ on $v_{2}$ of shape-selected events in 30--40\% centrality class in Pb--Pb collisions at $\sqrt{s_{\rm NN}}=\rm 5.02$ TeV. Right: Extracted CMW fraction as a function of centrality.} 
    \label{fig:ese}
    \end{figure}

    By implementing the ESE technique, one can estimate the fraction of the CMW signal~\cite{RefLCC3}. The left panel of Fig. \ref{fig:ese} presents the $\Delta \rm Int. \, Cov.$ as a function of $v_{2}$ for shaped-selected events in the 30--40\% centrality interval. The solid line represents a linear fit while the colored band indicates the confidence of the fit within $1 \sigma$ (68\%). The dependence of $\Delta \rm Int. \, Cov.$ changing with shaped $v_{2}$ as well as the finite intercept indicate that the background from LCC plays an important role in the CMW measurement. 
    
    The right panel of Fig. \ref{fig:ese} shows a moderate centrality dependence of the CMW fraction ($f_{\rm CMW}$), which is defined as
    \begin{equation}
        f_{\rm CMW} = \frac{b}{a \times \bar{v_2}+b},
    \end{equation}  
    where a and b are the slope and the intercept from the linear fit, respectively, illustrated in the left panel of Fig. \ref{fig:ese}. The total, centrality-independent systematical uncertainty is denoted by the shaded box on the edge of the line at zero. The fit over the 10--60\% centrality interval yields a value of $f_{\rm CMW} \sim 0.338 \pm 0.084 \, \rm (\rm stat.) \pm 0.198 \, \rm (\rm syst.)$. It can be seen that $f_{\rm CMW}$ is compatible with zero within uncertainties, implying that the CMW signal could be very small if it exits.
\section{Summary}
\setParDis
    The $r_{\Delta v_2}^{\rm Norm}$ and $r_{\Delta v_3}^{\rm Norm}$ of charged hadrons have been measured in Pb--Pb collisions at $\sqrt{s_{\rm NN}}=\rm 5.02$ TeV. The $v_{3}$ results are comparable to those of $v_{2}$ which points to a large background contribution to the CMW. The ESE technique was applied to disentangle the CMW signal from the LCC background. The dependence of $\Delta \rm Int. \, Cov.$ on $v_{2}$ also indicates that LCC dominates the CMW measurements. According to fit parameters, one can extract the $f_{\rm CMW}$ in different centrality classes. Averaging over the 10--60\% centrality interval, the measured $f_{\rm CMW}$ ($f_{\rm CMW} \sim 0.338 \, \pm  \, 0.084 \, (\rm stat.)  \, \pm  \, 0.198  (\rm syst.)$) is consistent with zero within experimental uncertainties in Pb--Pb collisions.
    
    This work was supported in part by the National Key Research and Development Program of China (No. 2018YFE0104600), the National Natural Science Foundation of China (Nos. 11975078, 12061141008) and the Shanghai Rising-Star Program (20QA1401500).

\nolinenumbers
\end{document}